\documentclass[%
 reprint,
 amsmath,amssymb,
 aps,
]{revtex4-2}
\usepackage{graphicx}%
\usepackage{dcolumn}%
\usepackage{bm}%
\usepackage{amsmath, amssymb}
\usepackage{tabularx}
\usepackage{hyperref}%
\usepackage[mathlines]{lineno}%
\usepackage{xcolor}
\usepackage[normalem]{ulem}
\definecolor{tangerine}{rgb}{0.944,0.522,0}
\definecolor{verde}{rgb}{0.,0.6,0}
\definecolor{rosso}{rgb}{0.9,0.0,0.2}
\definecolor{magenta}{rgb}{0.9,0.2,0.9}

\newif\ifhighlight

\newcommand{\highlight}{\highlighttrue}
\highlight
\newcommand{\editor}[2]{%
  \expandafter\newcommand\csname #1note\endcsname[1]{%
    \textcolor{#2}{(\textbf{#1note:} \textsc{##1})}}%
  \expandafter\newcommand\csname #1\endcsname[1]{%
    \ifhighlight\textcolor{#2}{##1} \else ##1\fi}%
  \expandafter\newcommand\csname #1cancel\endcsname[1]{%
    \ifhighlight\textcolor{#2}{\sout{##1}}\fi}%
  \expandafter\newcommand\csname #1change\endcsname[2]{%
    \ifhighlight\textcolor{#2}{\sout{##1} ##2}\else ##2\fi}%
  \newenvironment{#1text}{\ifhighlight\color{#2}\fi}{\color{black}}
}

\editor{FG}{rosso}
\editor{SC}{orange}
\editor{MC}{green}
\editor{WB}{magenta}

\newcommand{\resub}[1]{#1}

\begin{document}

\preprint{APS/123-QED}

\title{Adaptive energy reference for machine-learning models \\of the electronic density of states
}%

\author{Wei Bin How}
\affiliation{%
 Laboratory of Computational Science and Modeling, IMX,\\
 École Polytechnique Fédérale de Lausanne, Lausanne 1015, Switzerland
}%
\author{Sanggyu Chong}%
\affiliation{%
 Laboratory of Computational Science and Modeling, IMX,\\
 École Polytechnique Fédérale de Lausanne, Lausanne 1015, Switzerland
}%
\author{Federico Grasselli}
\altaffiliation[Current address:~]{%
\resub{Department of Physics, Informatics and Mathematics, University of Modena and Reggio Emilia, and
CNR-NANO S3---Istituto Nanoscienze, Modena 41125, Italy.} 
}
\author{Kevin K. Huguenin-Dumittan}
\affiliation{%
 Laboratory of Computational Science and Modeling, IMX,\\
 École Polytechnique Fédérale de Lausanne, Lausanne 1015, Switzerland
}%
\author{Michele Ceriotti}
\email{michele.ceriotti@epfl.ch}
\affiliation{%
 Laboratory of Computational Science and Modeling, IMX,\\
 École Polytechnique Fédérale de Lausanne, Lausanne 1015, Switzerland
}%

\begin{abstract}
The electronic density of states (DOS) provides information regarding the distribution of electronic energy levels in a material, and can be used to approximate its optical and electronic properties and therefore guide computational materials design.
Given its usefulness and relative simplicity, it has been one of the first electronic properties used as target for machine-learning approaches going beyond interatomic potentials. 
A subtle but important point, well appreciated in the condensed matter community but usually overlooked in the construction of data-driven models, is that for bulk configurations the absolute energy reference of single-particle energy levels is ill defined.
Only energy differences matter, and quantities derived from the DOS are typically independent of the absolute alignment.
We introduce an adaptive scheme that optimizes the energy reference of each structure as part of the training process, and show that it consistently improves the quality of machine-learning models compared to traditional choices of energy reference, for different classes of materials and different model architectures. 
On a practical level, we trace the improved performance to the ability of this self-aligning scheme to match the most prominent features in the DOS. 
More broadly, we believe that this paper highlights the importance of incorporating insights on the nature of the physical target into the definition of the architecture and of the appropriate figures of merit for machine-learning models, that translate in better transferability and overall performance. 
\end{abstract}

\maketitle

\section{\label{Introduction}Introduction} %

The electronic density of states (DOS) characterizes the distribution of available energy states for electrons and can be used to approximate many properties of a material \cite{toriyama_how_2022, ashcroft_solid_1976}, such as its electrical conductivity and optical properties, and can be used as a screening and inverse design tool for computational materials discovery. 
Given the good balance between accuracy and cost, the DOS is often computed using density functional theory (DFT), where the electrons are treated as non-interacting fermions and their interactions are modeled by a self-consistent mean-field potential. Even within this approximate treatment, the calculation of the DOS scales cubically with the number of electrons in the system \cite{chandrasekaran_solving_2019}. Hence, studying the DOS of large and complex systems is still prohibitively difficult.

Recently, machine learning (ML) has been shown to be a promising alternative approach to circumvent the expensive DFT calculations for large systems \cite{ceriotti_machine_2021, prezhdo_advancing_2020, noe_machine_2020, poltavsky_machine_2021, ramprasad_machine_2017, musil_physics-inspired_2021, glielmo_unsupervised_2021}. Over the past two decades, there have been significant developments in the integration of ML in computational materials science, ranging from the development of descriptors \cite{nigam_equivariant_2022,huguenin-dumittan_physics-inspired_2023,drautz_atomic_2019, bartok_representing_2013} to different model architectures \cite{schutt_schnet_2017, pozdnyakov2024smoothexactrotationalsymmetrization, lee_predicting_2023} to predict various material properties with high accuracy and low cost. Aside from the prediction of material properties, there has also been significant advancements in evaluating the reliability of these methods \cite{chong_robustness_2023, kailkhura_reliable_2019} and improving the efficiency of ML via dimensionality reduction of the feature space \cite{how_significance_2021, how_dimensionality_2022, kabylda_efficient_2023, zhu_principal_2022, lopanitsyna_modeling_2023}. 

Even though early efforts in this field focused on predicting the interatomic potentials \cite{behler_generalized_2007}, more recently there have been works geared toward modeling the electronic structure, including the DOS. Broderick and Rajan demonstrated that the prediction of the DOS can be done by decomposing it using principal component analysis and subsequently predicting the components using elemental descriptors \cite{broderick_eigenvalue_2011}. This approach was then further extended to metal alloys by Yeo \textit{et al}. and applied together with graph neural networks to metallic nanoparticles by Bang \textit{et al}. \cite{yeo_pattern_2019, bang_accelerated_2021}. Another approach was proposed by del Rio and Chandrasekeran \textit{et al}., using multilayer perceptrons (MLPs) to effectively learn and predict the DOS discretized on an energy grid \cite{del_rio_efficient_2020, chandrasekaran_solving_2019}. Similarly, this approach has also been extended to more complex models such as graph neural networks and attention networks \cite{fung_machine_2021, kong_density_2022}. Furthermore, Ben Mahmoud \textit{et al}. performed the prediction of the DOS, both on the discretized energy grid and the principal components, using kernel ridge regressors \cite{ben_mahmoud_learning_2020}.

One detail that has been often overlooked in previous works is the lack of an absolute electronic energy reference in the quantum mechanical calculations of infinite bulk systems \cite{zunger_first-principles_1978}. Typically, the energy reference is defined as the average Hartree potential in the cell \cite{blum_ab_2009} based on
\begin{align}
    V_{\text{H}} = \frac{1}{\Omega}\int_{cell} d^3\mathbf{r}\, V(\mathbf{r}).
\end{align}
where $V(\mathbf{r})$ represents the Hartree potential at point \textbf{r}, and $\Omega$ represents the volume of the unit cell. Due to the conditionally convergent nature of the potentials in an infinite bulk system, the average Hartree potential is usually ill defined\cite{kleinman_comment_1981}. 

This issue is very well recognized in condensed-matter modeling, and several approaches have been developed to alleviate its impact by defining an external energy reference within the system. One class of methods attempts to define the vacuum potential within the cell. This can be achieved by creating a surface, introducing sufficient amounts of empty space into the unit cell for the potential to converge to a constant value \cite{logsdail_bulk_2014}. For sufficiently porous systems, the vacuum potential can also be estimated as the spherical average of the electrostatic potential at the center of the pore \cite{butler_electronic_2014}.
An alternative approach is to build a cluster model of the lattice embedded in the potential of the periodic crystal \cite{scanlon_band_2013}. By virtue of being a cluster model, the calculation will include a vacuum region where the energy reference can be defined. 
Nevertheless, these methods only work well for specific system types and cannot be generalized to all bulk systems. Additionally, it is important to note that a true external energy reference for an infinite bulk system is still ill defined and these approaches only serve to provide a proxy for it. \resub{Due to the lack of an absolute energy reference,} only the quantities that are independent of the energy reference can be compared against experimental observables \cite{baldereschi_band_1988}.

\resub{Conversely, given that experimentally relevant quantities of interest are independent of the energy reference, the choice of such reference is arbitrary and inconsequential.} However, most ML methods do not treat the DOS in a way that is independent of the energy reference, but instead employ either the average Hartree potential, $V_\mathrm{H}$, or the Fermi level, $E_{\mathrm{F}}$, of the system as the energy reference of each structure in the dataset. Therefore, the energy reference chosen for each structure can have an impact on the performance of a ML model. For instance, an energy reference that is poorly defined across the dataset can lead to common spectral features being shifted to different energy ranges for each structure. This artificially increases the complexity of the dataset and can introduce discontinuities in the mapping between descriptors and targets that hamper model learning. 

In this paper, we present a ML framework that treats the energy reference of the DOS of \textit{each} structure in the dataset as an optimizable parameter. The framework is demonstrated on descriptors built on the smooth overlap of atomic positions (SOAP) power spectrum using both linear and nonlinear models. We showcase the effectiveness of this approach in improving the performance of machine learning predictions of the DOS and its derived quantities. Four datasets of varying complexity and elemental compositions are used to assess the performance of this framework. For each dataset, the framework is compared against the other commonly used energy references, $V_\mathrm{H}$ and $E_{\mathrm{F}}$.  We show that treating the energy reference of the DOS as an optimizable parameter consistently leads to the best performing models for both the DOS and the observables that can be derived from it, and rationalize these results in terms of the alignment of the main spectral features of the DOS. Furthermore, we also highlight the enhanced transferability of the adaptive-reference models to larger system sizes. This framework can improve the generalizability of ML methods for the prediction of the DOS, especially on datasets where common spectral patterns are not well aligned when employing conventional energy references.  

\section{Methods}
Although experimentally accessible quantities derived from the DOS are independent of the energy reference, ML models are typically trained and evaluated in a way that does depend on the relative alignment of the energy reference chosen for the dataset. This is seen in the mean squared error (MSE) loss function that is typically employed for ML DOS:
\begin{align}
    \label{Typical_Loss}L(\bm{W}) &= \frac{1}{N} \sum_A \int dE \Biggl (\mathrm{DOS}^Q_A(E)-\mathrm{DOS}^{{\bm{W}}}_{A}(E) \Biggr)^2.
\end{align}
In this loss function, $N$ represents the number of structures in the dataset, $\bm{W}$ represents the parameters of the model, $\mathrm{DOS}^Q_A(E)$ represents the DOS obtained from quantum chemical calculations for structure~$A$, and $\mathrm{DOS}_{A}^{{\bm{W}}}(E)$ represents the DOS predicted by the model with parameters $\bm{W}$. Here, changing the energy reference of $\mathrm{DOS}^Q_A(E)$ by replacing $E$ with $E + \Delta$, where $\Delta$ parameterizes the changes in the energy reference, would result in a different loss value.

One solution to this would be to redefine the training and evaluation metrics to be independent of the energy reference. This can be achieved by reformulating the MSE loss such that the predicted DOS is evaluated on the energy reference that minimizes the loss:
\begin{align}
    \label{Aligned_Loss}L(\bm{W}) = \frac{1}{N} \sum_A  \min_{\Delta_A} \Biggl [\int dE \Biggl (&\mathrm{DOS}^Q_A(E + \Delta_A) \\ 
    & -\mathrm{DOS}^{{\bm{W}}}_{A}(E) \Biggr)^2 \Biggr] \nonumber.
\end{align}
In this revised MSE, $\Delta_A$ is a parameter that shifts the energy reference of $\mathrm{DOS}^Q_A$, and is chosen to be the value that minimizes the MSE for the DOS of structure~$A$ predicted by the model. 

\begin{figure*}
\centering
\includegraphics[trim=1.5cm 6cm 0.5cm 6cm, clip, width=1\linewidth]{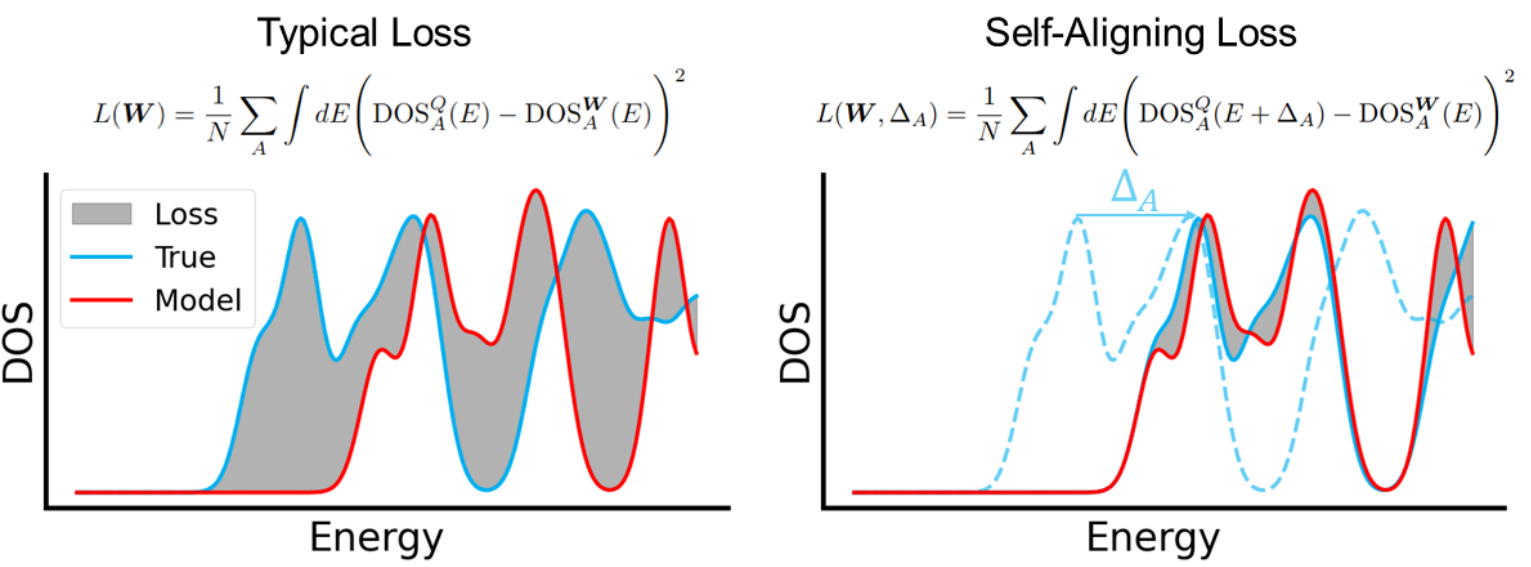}
\caption{\label{fig:MSEfig} A schematic that highlights the difference between the standard loss and the self-aligning loss in training ML models for DOS predictions.}
\end{figure*}

Although Eq.~\eqref{Aligned_Loss} is a better metric to evaluate the quality of trained models, implementing it as a loss to drive the training process results in significantly longer training times, as seen in Sec. I of the Supplemental Material. In view of this, our approach for DOS learning instead treats $\Delta_A$ of the structures in the training dataset as a parameter to be optimized alongside the model parameters during model training, corresponding to a self-aligning loss: 
\begin{align}
    \label{A_Loss}L(\bm{W}, \Delta_A) &= \frac{1}{N} \sum_A \int dE \Biggl (\mathrm{DOS}^Q_A(E + \Delta_A) \\
    & -\mathrm{DOS}^{{\bm{W}}}_{A}(E) \Biggr)^2 \Biggr] \nonumber,
\end{align}
\resub{where $\Delta_A$ is an optimizable parameter that shifts the energy reference of $\mathrm{DOS}^Q_A$.} Fig. \ref{fig:MSEfig} shows a schematic of the effect of optimizing $\Delta_A$ for a self-aligning loss.
By introducing $\Delta_A$ as an optimizable parameter, the model is free to adjust the energy reference while training. This way, the loss of the model would not be artificially inflated by a choice of energy reference that has poor alignment between the structures in the dataset, which is not physically relevant. As a result, the weights of the model would be more responsive to the loss induced by the physically meaningful spectral details of the DOS. We show in the Supplemental Material that the joint optimization of $\Delta_A$ and $\bm{W}$ through the loss \eqref{A_Loss} achieves similar results to the optimization of $\bm{W}$ using the loss \eqref{Aligned_Loss} at much shorter training times. Additionally, the joint optimization framework can be easily integrated into existing DOS ML workflows via a simple modification of the model training loop (Sec. \ref{Model Training}). The remainder of this section discusses the structure of the model and the construction of the ML targets.

\subsection{Model details}\label{Model Training}
We employ a locality ansatz where the global DOS of a structure $A$ is decomposed into contributions from each atomic environment, $A_i$, where i runs over all atoms in structure $A$,
\begin{align}
    \label{LDOS decomposition}\mathrm{DOS}_A(E) = \frac{1}{N_A}\sum_{A_{i} \in A} \mathrm{LDOS}_{A_{i}}(E),
\end{align}
where $N_A$ represents the total number of atomic environments in structure $A$ and $\mathrm{LDOS}_{A_{i}}$ represents the local contribution of environment $A_i$ to the global DOS. Additionally, the models are constructed such that they predict the DOS at different energy values on a grid in a multitarget regression manner. 

To account for differences in system sizes across the dataset, \resub{we always plot and assess the accuracy of the DOS after normalizing it by the number of atoms in the structure. However, we do not explicitly enforce a normalization of the DOS predictions to the number of available energy levels, because we usually avoid predicting the (inherently inaccurate) region corresponding to the highest energy levels. We initialize the DOS alignment --- and use as baseline alignment scheme --- based on the Fermi level $E_{\mathrm{F}}$, as it usually provides much better results than the mean Hartree potential.}Furthermore, to only capture the relative energy references across the structures in the dataset, the mean of $\Delta_A$ is subtracted out such that the mean change in energy reference across the dataset is 0. For models trained without optimizing the energy reference, the loss was defined as in Eq. \eqref{Typical_Loss}.

In this paper, our framework was incorporated in both linear models and MLPs, using the SOAP power spectrum as input features. The MLPs have one hidden layer, with a number of neurons equal to the number of input features. Specific details about the SOAP power spectrum, model architecture and training can be found in Sec. III of the Supplemental Material.
\subsection{Constructing the DOS}\label{DOS construction}

To construct the DOS for a bulk structure from a finite number of k points, one can apply Gaussian broadening to the eigenenergies,
\begin{align}
     \label{DOS}\mathrm{DOS}_A^{\delta}(E) &= \frac{1}{N_A}\frac{2}{N_k} \sum_{n \in \mathrm{bands}}\sum_\textbf{k}g(E-\epsilon_n(\textbf{k}) -\delta_A) \\
     g(x) &= \frac{1}{\sqrt{2\pi\sigma^2}}e^{-\frac{x^2}{2\sigma^2}},
\end{align}
where $\delta \in \{ V_\mathrm{H}, E_\mathrm{F}\}$, indicates the fixed energy reference used to define the dataset, $N_k$ indicates the number of k points from which the eigenenergies were computed from, $\epsilon_n(\textbf{k})$ represents the eigenenergy of each band, $n$, at different k points, and $\sigma$ represents the Gaussian broadening parameter. The DOS is normalized with respect to $N_A$ to allow for meaningful comparisons between different systems with different number of atoms.

Since the energy reference for each structure is defined to be its cell average Hartree potential, the eigenenergies, $\epsilon_n(\textbf{k})$, are also referenced to the cell average Hartree potential by default. To obtain the DOS referenced to $V_\mathrm{H}$ or the Fermi level, the energy reference $\delta$ is set to either zero or the Fermi level with respect to $V_\mathrm{H}$, respectively. The Fermi level, $E_{\mathrm{F}}$, relative to $V_\mathrm{H}$ is defined such that:

\begin{align}
	\label{Ef}\mathrm{n_{val}}(A) &= \int_{-\infty}^{E_\mathrm{F}^A} dE \, \, \mathrm{DOS}^{V_\mathrm{H}}_A(E)
\end{align}
where $\mathrm{n_{val}(}A\mathrm{)}$ indicates the average number of valence electrons per atom used in the DFT calculation of system $A$, and $E_\mathrm{F}^A$ is the Fermi level of structure $A$ relative to $V_\mathrm{H}$.

As the DOS is represented on a discretized energy grid, calculating the change, and the derivative, of the DOS with respect to a change in the energy reference would be computationally expensive as one would have to recalculate Eq.~\eqref{DOS} again, for a large number of structures and eigenenergies. To tackle this, cubic Hermite splines \cite{catmull_class_1974} are built on the DOS, where the DOS function is represented piecewise by a third-degree polynomial that is defined by the value and derivatives at the end points of the interval. Given that $\Delta_A$ in Eq. \eqref{Aligned_Loss} and Eq. \eqref{A_Loss} is always applied on $\mathrm{DOS}^Q_A$ instead of $\mathrm{DOS}^{\bm{W}}_A$, the splines only need to be built once before model training.

Additionally, to evaluate the performance of the models on derived quantities of the DOS that are independent of the energy reference, secondary observables such as DOS($E_{\mathrm{F}}$) and the distribution of excitations, $X(\theta)$, are also calculated for each structure. The equation for the distribution of excitations is defined as follows \cite{ben_mahmoud_learning_2020}:

\begin{align}
	\label{AOFD} X_A(\theta) &= \iint dE dE' \, \mathrm{DOS}^\delta_A(E)f_{\mathrm{FD}}(E - E_\mathrm{F}^A)_{T = 0} \\ &\mathrm{DOS}^\delta_A(E')(1 - f_{\mathrm{FD}}(E' - E_\mathrm{F}^A)_{T = 0})\delta(E-E'-\theta) \nonumber
\end{align}
It can be observed that $X_A(\theta)$ is independent of the energy reference since the Fermi level of the system, $E_\mathrm{F}^A$, acts as an independent point of reference. In this paper, the excitation energy, $\theta$, is defined on an energy grid ranging from 0 to 2 eV, with a 0.05 eV grid spacing.

\begin{figure*}
\centering
\includegraphics[width=1\linewidth]{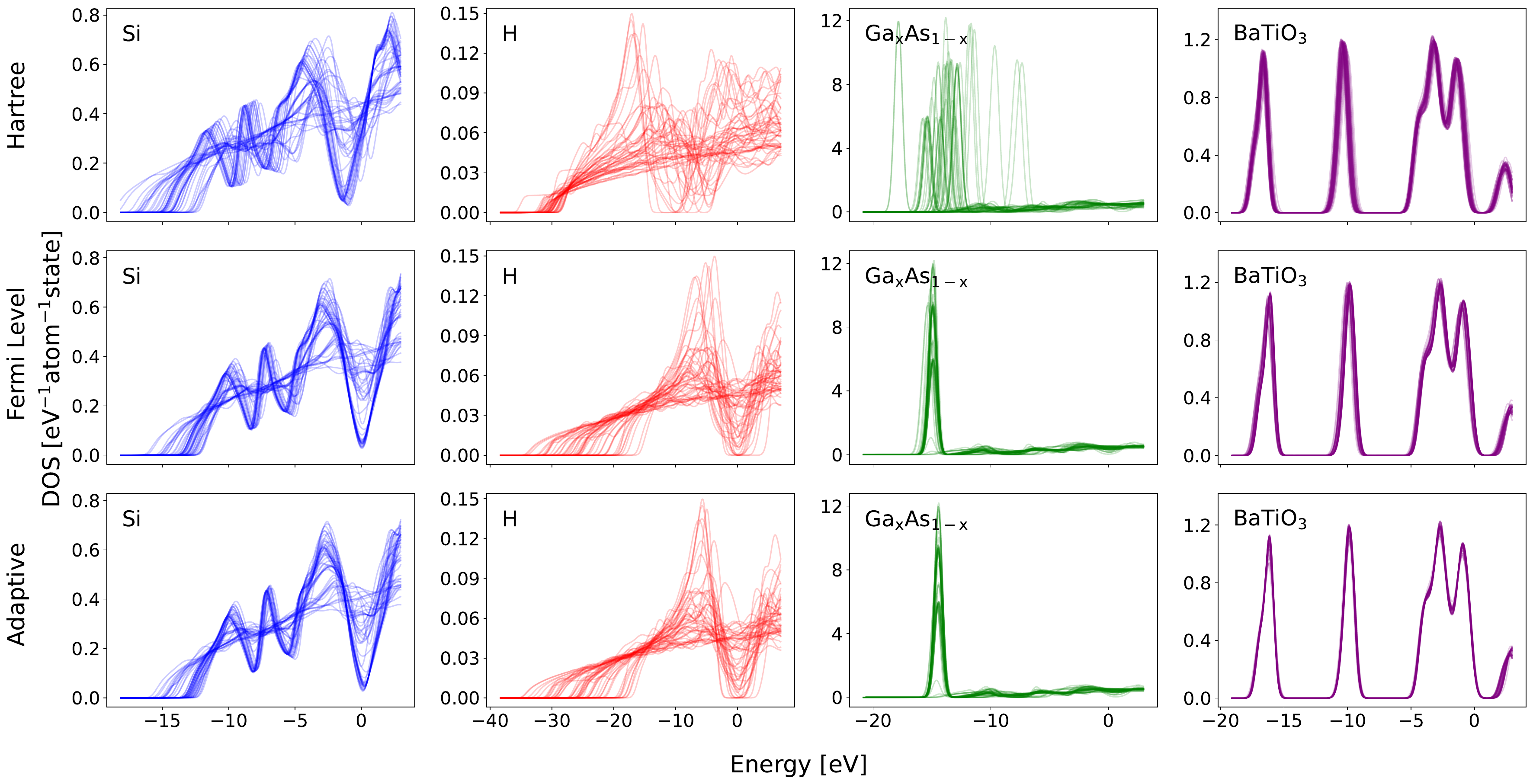} 
\caption{\label{fig:Intro_Plot} DOS overlay plots of each dataset with different energy references is shown in each panel. Each overlay plot contains 50 different lines, corresponding to the DOS of 50 different structures in the dataset indicated in each panel. \resub{Each row depicts the DOS under a different energy reference. In the first row, the energy reference used is the average Hartree potential ($V_\mathrm{H}$), followed by the Fermi level ($E_{\mathrm{F}}$) in the second, and the adaptive energy reference in the third. The curves relative to $V_\mathrm{H}$ have been rigidly shifted by a global value to visually align with those computed using $E_{\mathrm{F}}$. Due to the significant band gap in $\mathrm{BaTiO_3}$, the highest occupied molecular orbital (HOMO) was used in place of $E_{\mathrm{F}}$ for the initial alignment and the visualization.}}
\end{figure*}

\subsection{Datasets}{\label{Datasets}}

A total of four datasets were used to evaluate the effectiveness of the joint optimization scheme. The first dataset comprises 927 silicon structures obtained from Bart$\mathrm{\acute{o}}$k \textit{et al}. \cite{bartok_machine_2018}, spanning across four different bulk phases: diamond, beta-tin, liquid and amorphous bulk, and surfaces. DFT calculations of the structures were performed using FHI-aims \cite{blum_ab_2009} with the parameters employed by Ben Mahmoud \textit{et al}. \cite{ben_mahmoud_learning_2020}. Additionally, 211 diamond defect structures of two different sizes, 63 and 215 atoms, were also obtained from Bart$\mathrm{\acute{o}}$k \textit{et al}. \cite{bartok_machine_2018} for evaluating the transferability of the models to larger system sizes. The second dataset is adapted from Cheng \textit{et al}. \cite{cheng_evidence_2020} and is composed of bulk hydrogen configurations at high pressures, including a phase transition between solid and liquid hydrogen. 20,000 structures were subselected from the entire dataset. Of the 20,000 structures, 5000 structures were obtained from ab initio molecular dynamics, and 2143 structures were selected from each of the seven iterations of a random structure search simulation\cite{pick-need11jpcm} using farthest point sampling. DFT calculations of the structures were performed using Quantum Espresso \cite{giannozzi_advanced_2017, giannozzi_q_2020, giannozzi_quantum_2009} with the parameters employed by Ben Mahmoud \textit{et al}. \cite{ben_mahmoud_predicting_2022}. 
\begin{table}
\begin{ruledtabular}
\begin{tabular}{ccccc}
 \multicolumn{5}{c}{\resub{Standard deviations of targets}}\\
&\resub{DOS$^{V_\mathrm{H}}$}&\resub{DOS$^{E_{\mathrm{F}}}$}&\resub{DOS$(E_{\mathrm{F}})$}&\resub{$X(\theta)$}\\ \hline
 \resub{Si}&\resub{0.367}&\resub{0.284}&\resub{0.144}&\resub{0.0880} \\
 \resub{H}&\resub{0.104}&\resub{0.0711}&\resub{0.0187}&\resub{0.00200} \\
 \resub{$\mathrm{BaTiO_3}$}&\resub{0.318}&\resub{0.244}&\resub{0.000139}&\resub{0.00251} \\
 \resub{Ga$_x$As$_{1-x}$}&\resub{5.67}&\resub{2.69}&\resub{0.124}&\resub{0.116} \\
\end{tabular}
\end{ruledtabular}
\caption{\label{tab:StdDevs}\resub{Standard deviation for the targets across datasets. DOS$^{V_\mathrm{H}}$ and DOS$^{E_{\mathrm{F}}}$ refers to the DOS constructed with the Hartree potential and Fermi level energy reference respectively. Due to the significant bandgap present in the structures of $\mathrm{BaTiO_3}$, the highest occupied molecular orbital (HOMO) was used in place of $E_{\mathrm{F}}$ to define the Fermi level energy reference. Note the minuscule variability of  DOS$(E_{\mathrm{F}})$ for $\mathrm{BaTiO_3}$, compatible with its nature of a large band-gap insulator. The very large variability of DOS$^{V_\mathrm{H}}$ is very high for Ga$_x$As$_{1-x}$, due to the poor alignment of the core levels when using $V_\mathrm{H}$ as the reference.}} 
\end{table}
The third dataset involves bulk $\mathrm{BaTiO_3}$ structures, a perovskite crystal which exhibits strong ferroelectric properties and has technological relevance \cite{karvounis_barium_2020}. Four phases, rhombohedral, orthorhombic, tetragonal, and cubic, are included in the dataset. The dataset comprises 840 structures, with 210 structures in each phase and is obtained from the NpT subset of the $\mathrm{BaTiO_3}$ dataset built by Gigli \textit{et al} \cite{gigli_thermodynamics_2022}.
The last dataset is that of a $\mathrm{Ga}_x\mathrm{As}_{1-x}$ binary system over a wide range of temperatures, pressures, and stoichiometry in order to accurately represent the changes in the electronic behavior of the system across the binary phase diagram. This is particularly important as GaAs is a semiconductor with excellent electronic and optical properties, lending itself applications in solar cells and photonics \cite{verrinder_gallium_2022, li_widebandgap_2020}.
The dataset is obtained from Imbalzano and Ceriotti \cite{imbalzano_modeling_2021} and is composed of crystalline structures, liquid phases, and various interfaces. The dataset is then recomputed at a higher level of theory to account for van der Waals interactions and a larger number of electronic bands. The computational parameters for all the datasets used are detailed in Sec. IV of the Supplemental Material.

With the exception of the hydrogen dataset, the predicted DOS energy range spans from 1.5 eV below the lowest eigenenergy to 3 eV above the maximum Fermi level in the dataset, while the splines are defined up to 6 eV above the maximum Fermi level to enable shifts beyond the predicted DOS energy range. Meanwhile for the hydrogen dataset, due to high $\Delta_A$ values and high predicted $E_{\mathrm{F}}$, the DOS energy range instead spans up to 7 eV above the maximum Fermi level and the splines are defined up to 12 eV above the maximum Fermi level. For all datasets, both the splines and DOS energy range are discretized on a grid with a spacing of 0.05 eV. \resub{The standard deviations of the targets for each dataset are shown in Table. \ref{tab:StdDevs}, providing a scale to gauge the accuracy of predictions relative to the intrinsic variability of the target values.}

Each dataset is split randomly in a 7:1:2 ratio to give the training, validation and test set, respectively. Smaller subsets, taken at 25\%, 50\% and 75\% of the training and validation data, were also used to train the model to construct learning curves.

\section{Results}

\begin{table*}
\begin{ruledtabular}
\begin{tabular}{ccccccc}
 &\multicolumn{3}{c}{Linear models test $\mathrm{RMSE}^{\mathrm{DOS}}$}&\multicolumn{3}{c}{Multilayer perceptrons test $\mathrm{RMSE}^{\mathrm{DOS}}$}\\
  &\multicolumn{3}{c}{$[\mathrm{eV^{-0.5}atom^{-1}state}]$}&\multicolumn{3}{c}{$[\mathrm{eV^{-0.5}atom^{-1}state}]$}\\
 Dataset&$V_\mathrm{H}$&$E_{\mathrm{F}}$&Adaptive
&$V_\mathrm{H}$&$E_{\mathrm{F}}$&Adaptive\\ \hline
 Si&0.0578&0.0428&0.0399&0.0409&0.0332&0.0299 \\
 H&0.0332&0.0292&0.0268&0.0300&0.0256&0.0227 \\
 $\mathrm{BaTiO_3}$&0.0681&0.0267&0.0226&0.0288&0.0246&0.0182 \\
 Ga$_x$As$_{1-x}$&&0.203~&0.0993&&0.194~&0.0969 \\
\end{tabular}
\end{ruledtabular}
\caption{\label{tab:DOSRMSE} DOS RMSE of models trained on different datasets and different energy references. The DOS RMSE is evaluated on the test set and has units of $\mathrm{eV^{-0.5}atom^{-1}state}$. For Ga$_x$As$_{1-x}$, there were no models trained using the cell average Hartree Potential as the energy reference due to the wide range of Fermi levels in the dataset. Additionally, due to the significant bandgap present in the structures of $\mathrm{BaTiO_3}$ the highest occupied molecular orbital (HOMO) was used in place of $E_{\mathrm{F}}$ to define the energy reference.}
\end{table*}
\begin{table*}
\begin{ruledtabular}
\begin{tabular}{ccccccc}
 &\multicolumn{3}{c}{Multilayer perceptrons $\mathrm{RMSE^{\mathrm{DOS}(E_F)}}$}&\multicolumn{3}{c}{Multilayer perceptrons $\mathrm{RMSE}^X$}\\
 &\multicolumn{3}{c}{$[\mathrm{eV^{-1}atom^{-1}state}]$}&\multicolumn{3}{c}{$[\mathrm{eV^{-0.5}atom^{-2}state^2}]$}\\
 Dataset&$V_\mathrm{H}$&$E_{\mathrm{F}}$&Adaptive
&$V_\mathrm{H}$&$E_{\mathrm{F}}$&Adaptive\\ \hline
 Si&0.0235&0.0236&0.0201&0.00702&0.00683&0.00614 \\
 H&0.00944&0.00994&0.00796&0.000857&0.000979&0.000843 \\
 $\mathrm{BaTiO_3}$&0.0953&0.141&0.0629&0.0207&0.0322&0.0123 \\
 Ga$_x$As$_{1-x}$&&0.0356&0.0362&&0.0254&0.0137 \\
\end{tabular}
\end{ruledtabular}
\caption{\label{tab:SecondaryRMSE} RMSE on secondary quantities for MLP models trained on different datasets and different energy references. All RMSEs are evaluated on the test set. The RMSE for $\mathrm{DOS}(E_{\mathrm{F}})$ has units of $\mathrm{eV^{-1}atom^{-1}state}$ and that of $X$ has units of $\mathrm{eV^{-0.5}atom^{-2}state^2}$. For Ga$_x$As$_{1-x}$, there were no models trained using the cell average Hartree potential as the energy reference due to the wide range of Fermi levels in the dataset. Additionally, due to the significant bandgap present in the structures of $\mathrm{BaTiO_3}$ the HOMO was used in place of $E_{\mathrm{F}}$ to define the energy reference.}
\end{table*}

\subsection{Evaluation of model performance}\label{Model Evaluation}
To assess the impact of the choice of energy reference, we evaluate the performance of each model on the DOS and its derived quantities using the root mean squared error (RMSE), defined as the square root of the corresponding MSE value, on the test set. To evaluate each model in a way that is independent of the energy reference, the MSE is calculated using an optimal alignment strategy, analogous to the definition in Eq. \eqref{Aligned_Loss}:
\begin{align}
    \label{Test RMSE}\mathrm{MSE}^{\mathrm{DOS}}(\bm{W}) &= \frac{1}{N} \sum_A  \min_{\Delta_A} \Biggl [\int dE \Biggl (\mathrm{DOS}^Q_A(E + \Delta_A) \\ 
    &-\frac{1}{N_A} \sum_{A_{i} \in A} \mathrm{LDOS}_{A_{i}}^{{\bm{W}}}(E) \Biggr )^2\Biggr]   \nonumber.
\end{align}
The variables follow the same conventions as those used in Eq.~\eqref{Aligned_Loss}. Additionally, $\mathrm{LDOS}_{A_{i}}^{{\bm{W}}}$ refers to the predicted atomic contribution to the DOS of atomic environment $A_i$, as in Eq. \eqref{LDOS decomposition}. For diagnostic purposes, we also use an energy-resolved MSE computed using the optimal alignment for each structure,
\begin{align}
    \label{Energywise RMSE}\mathrm{MSE^{DOS}}(\bm{W}, E) &= \frac{1}{N} \sum_A \Biggl (\mathrm{DOS}^Q_A(E + \Delta_A) \\
    &\left.-\frac{1}{N_A} \sum_{A_{i} \in A} \mathrm{LDOS}_{A_{i}}^{\bm{W}}(E) \right )^2.   \nonumber
\end{align}

For the derived quantities that are independent of the energy reference, namely DOS($E_{\mathrm{F}}$) and $X(\theta)$, the quantities are derived from the predicted DOS and compared against that from quantum chemical calculations. As DOS($E_{\mathrm{F}}$) is a scalar quantity, the MSE can be defined as
\begin{align}
    \label{DOSEf RMSE} \mathrm{MSE}^{\mathrm{DOS}(E_{\mathrm{F}})}(\bm{W}) &= \frac{1}{N}\sum_A( \Tilde{y}_A - y_A)^2,
\end{align}
where $\Tilde{y}_A$ represents the DOS($E_{\mathrm{F}}$) derived from the predicted DOS of structure $A$ and $y$ represents the DOS($E_{\mathrm{F}}$) derived from the corresponding quantum chemical calculations. Although $\Tilde{y}_A$ may represent only one point on the predicted DOS spectrum, it depends indirectly on the entire DOS, up to the Fermi level due to the normalization condition in Eq. \eqref{Ef}.

Since $X_A(\theta)$ is a function of the excitation energy, $\theta$, the errors have to be integrated across $\theta$ as defined in the following,
\begin{align}
    \label{AOFD error} \mathrm{MSE}^X(\bm{W}) &= \frac{1}{N}\sum_A \int  d\theta\biggl ( \Tilde{X}_A(\theta) - X_A(\theta) \biggr )^2.
\end{align}
where, $\Tilde{X}_A$ represents the distribution of excitations obtained from the predicted DOS of structure $A$ and $X_A$ is derived from the corresponding quantum chemical calculation.

\begin{figure*}
\centering
\includegraphics[width=1\linewidth]{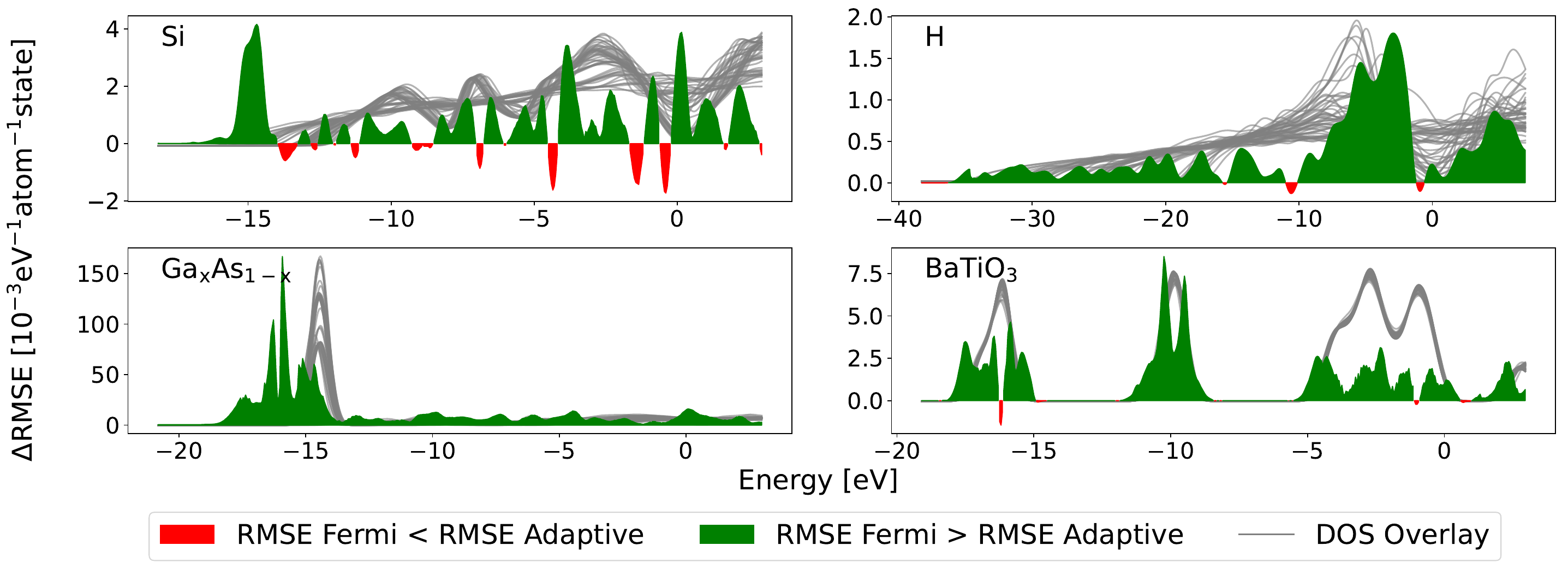} 
\caption{\label{fig:energywiseerrors} Difference in energy-resolved RMSE for the MLP models trained using the Fermi level versus the adaptive energy reference for the dataset. Green regions indicate where the model trained on the adaptive energy reference perform better, while red regions indicate that the Fermi level energy reference perform better. \resub{The gray lines show the DOS overlay of the adaptive energy reference, shown in the third row of Fig. \ref{fig:Intro_Plot}, in the same plot. It is important to note that the gray lines do not share the same scale as the green and red curves and they are only plotted as a visual cue to relate the change in error to the main DOS features}. It can be seen that models trained on the adaptive energy reference outperform those referenced to the Fermi level across the majority of the energy range. \resub{Additionally, for Ga$_x$As$_{1-x}$ and $\mathrm{BaTiO_3}$, it can be observed that spectral alignment coincides with the regions of largest improvement.}}
\end{figure*}

With the exception of the Ga$_x$As$_{1-x}$ dataset, three different energy references, $V_\mathrm{H}$, $E_{\mathrm{F}}$, and the optimizable energy reference, were employed for each dataset and their results were compared against one another. For the Ga$_x$As$_{1-x}$ dataset, the $V_\mathrm{H}$ energy reference was not employed due to the wide range of Fermi levels relative to $V_\mathrm{H}$, at around 24 eV. To truncate the DOS energy range at 3 eV above the maximum Fermi level, the DOS energy range would extend up to 27 eV above the Fermi level for certain structures. Such a wide energy range exceeds the maximum eigenenergy calculated for some structures resulting in an unphysical behavior, whereby the DOS decays quickly to zero at high energy levels for those structures due to a limitation in the number of bands calculated. Thus, aside from the Ga$_x$As$_{1-x}$ dataset, a total of six different models were trained for each dataset and their performance on the DOS RMSE on the test set are shown in Table \ref{tab:DOSRMSE}. As the adaptive reference provides the best performance on the secondary quantities for both MLPs and linear models, only the results for the MLPs are shown in Table \ref{tab:SecondaryRMSE} while that of the linear models are shown in Sec. V of the Supplemental Material. 

\subsection{Discussion}\label{Discussion}
\resub{Results show that} MLPs consistently perform better than linear models in predicting the DOS on the test set (Table \ref{tab:DOSRMSE}), which is expected considering that MLPs have more parameters and can also capture nonlinear relationships between the features and the DOS. As for the choice of energy reference, the Fermi level usually leads to smaller RMSE than the average Hartree potential. \resub{The adaptive energy alignment approach further improves the accuracy of the ML models, where the degree of improvement can vary significantly depending on the dataset of interest}. For the silicon and hydrogen datasets, the improvement in RMSE is rather small at roughly $10\%$ compared to the Fermi level alignment, with relatively larger improvements for MLPs compared to linear models. Meanwhile, for Ga$_x$As$_{1-x}$, the improvements against Fermi level alignment are roughly $50\%$ for both types of models, while it is $15\%$ and $25\%$ for linear and MLPs respectively on the $\mathrm{BaTiO_3}$ dataset.

\resub{The enhancements in the ML model performance with the adaptive energy alignment scheme can be attributed to the improved matching of the position of prominent spectral features. This effect is most pronounced for the Ga$_x$As$_{1-x}$ and $\mathrm{BaTiO_3}$ datasets (see Fig. \ref{fig:Intro_Plot}), where the lowest lying 3d band of Ga become best aligned under the adaptive alignment scheme in the Ga$_x$As$_{1-x}$ dataset, and similarly the first two peaks in the case of the $\mathrm{BaTiO_3}$ dataset. As optimal alignment between such recurring energy state in the dataset is achieved, ML models are allowed to focus primarily on accurately resolving the finer features and patterns in the DOS without being burdened with the complex task of correctly placing these commonly appearing states at different energy levels, which may depend on global quantities such as the concentration of defects in a structure, and are therefore not straightforwardly associated with the local structural motifs upon which the ML model is based.}

\begin{figure*}
\centering
\includegraphics[width=1\linewidth]{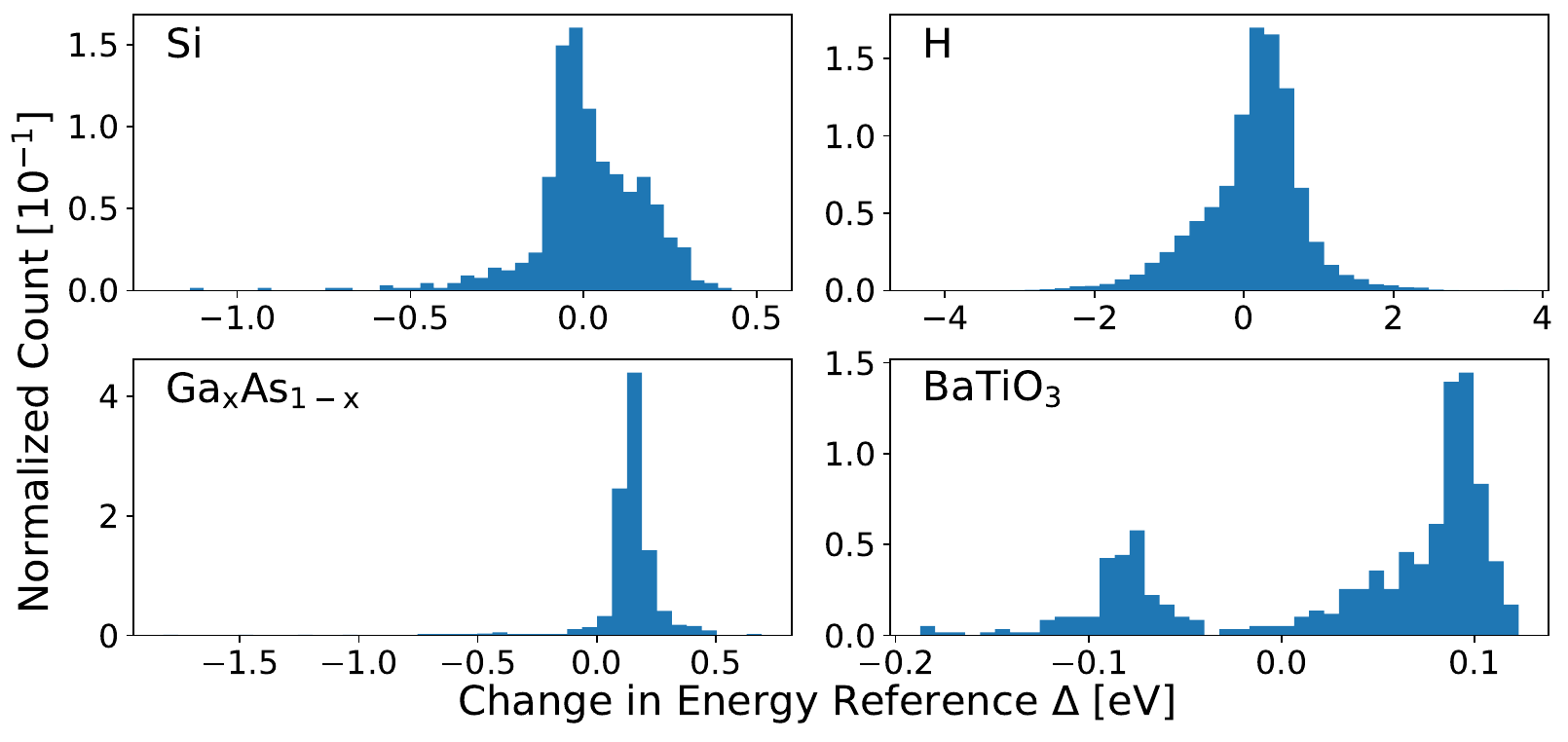} 
\caption{\label{fig:histogram} Histogram of $\Delta_A$ for the training structures for each dataset for MLPs. The histograms are normalized with respect to the sizes of the training set to make the sum of all bins equal to 1.}
\end{figure*}

Similar trends can be generally observed for the secondary quantities derived from the DOS predictions on the test set. Secondary quantities derived from the predictions of models trained on the adaptive energy reference typically performs the best. The degree of improvement is less significant for silicon and hydrogen while it is usually more pronounced for Ga$_x$As$_{1-x}$ and $\mathrm{BaTiO_3}$. Although there were no significant changes in the performance of DOS($E_{\mathrm{F}})$ for Ga$_x$As$_{1-x}$, improvements of roughly 15\% can be observed for silicon and hydrogen and $\mathrm{BaTiO_3}$ exhibits more substantial improvements --- \resub{ a factor of 2 relative to the accuracy of a model using a Fermi level alignment, and 34\% relative to using $V_\mathrm{H}$, which in this case is the best fixed reference. 
Note that the error for DOS($E_{\mathrm{F}})$ is very large in the case of $\mathrm{BaTiO_3}$, especially in comparison to the minuscule variability of this quantity in the dataset (see Table~\ref{tab:StdDevs}). 
This is due to the fact that --- barium titanate being a large band gap insulator --- small errors in the DOS values in the occupied bands can dramatically shift the Fermi level, moving it toward the states below or above the gap.  }
Meanwhile, for the predictions of $X(\theta)$, there was a negligible change in performance for hydrogen but there were improvements of 10\% of silicon and significantly more improvements for Ga$_x$As$_{1-x}$ and  $\mathrm{BaTiO_3}$ at 41\% and 46\%, respectively.

To investigate the model performance at different energy channels, the DOS RMSE was compared energywise between MLPs trained using the $E_{\mathrm{F}}$ and self-aligned energy reference. To calculate the energy-resolved RMSE, Eq. \eqref{Energywise RMSE} was used. The results are shown in Fig. \ref{fig:energywiseerrors} where the vertical axis denotes the change in RMSE, calculated by subtracting the RMSE on the adaptive reference from the RMSE on the Fermi level reference. 
From the lower quadrants, it can be seen that the improvements in DOS performance when using the adaptive energy reference are most prominent at the energy ranges of common spectral features. This corroborates the previous statement, whereby the misalignment of these features hinders the model performance by further complicating the learning problem.

Additionally, in Fig. \ref{fig:histogram}, the histograms of the $\Delta_A$ values for the training structures in each dataset also support this interpretation. The two distinct peaks at positive and negative $\Delta_A$ for $\mathrm{BaTiO_3}$ are indicative of the multimodal nature of the dataset that contains structures sampled from several distinct ferroelectric phases. 
These observations suggest that optimizing the energy reference to improve model performance results in the alignment of common spectral patterns in the dataset. Hence, this accounts for the disparity in improvement in model performance for different datasets, as the effectiveness of this framework will depend on the degree of similarity in the DOS spectra within the dataset, and on whether one of the common choices of reference leads to substantial shifts of these features in different structures. 

\resub{Finally, we consider effects beyond improvements in accuracy and assess the transferability of the ML models under different energy alignment schemes.
For this, we target vacancy defects in bulk silicon and retrain the models on an extended Si dataset which also includes 100 small vacancy structures with 63 atoms, randomly split into training and validation sets in a 7:1 ratio. The models were then evaluated on 111 large vacancy structures with 215 atoms, and the results are shown in Table \ref{tab:DefectRMSE}}. From the table, it can be seen that using the adaptive energy reference results in significantly better transferability to larger system sizes compared to the best performing fixed reference \resub{for each target, with a reduction of the RMSE by 38\%, 23\% and 59\% compared to the better one of the two pre-existing alignment schemes for DOS, $\mathrm{DOS}(E_{\mathrm{F}})$, and $X(\theta)$, respectively.
This example showcases the better resilience of a self-aligning model with respect to factors, such as defect concentration, that affect the internal energy references while being poorly reflected in the local descriptors that underlie most ML models.
}

\begin{table} [h]
\begin{ruledtabular}
\begin{tabular}{cccc}
 &\multicolumn{3}{c}{Multilayer perceptrons performance}\\
 RMSE&$V_\mathrm{H}$&$E_{\mathrm{F}}$&Adaptive\\ \hline
 DOS&0.0354&0.0343&0.0212\\
 DOS$(E_{\mathrm{F}})$&0.0141&0.0331&0.0108\\
 $X(\theta)$&0.00424&0.00392&0.00159\\
\end{tabular}
\end{ruledtabular}
\caption{\label{tab:DefectRMSE} RMSE on the DOS and secondary quantities on the large (215-atoms) vacancy structures for MLP models trained on the silicon dataset, extended by including only 63-atoms vacancy structures. The units are consistent with the previous tables, with $\mathrm{eV^{-0.5}atom^{-1}state}$,  and $\mathrm{eV^{-1}atom^{-1}state}$, and $\mathrm{eV^{-0.5}atom^{-2}state^2}$ for the units of the RMSE of the DOS, $\mathrm{DOS}(E_{\mathrm{F}})$, $X(\theta)$ respectively.}
\end{table}

\section{Conclusions}
To conclude, we propose an improved DOS learning framework that optimizes the energy reference of each structure. We compared the performance of this framework against two conventional energy alignment approaches: the mean Hartree potential and the Fermi level. 
Across four datasets, models trained under the adaptive framework consistently outperform others in terms of the DOS and its derived quantities. Additionally, we provide some justification behind the optimization process. Energy ranges in the DOS that experience the greatest improvement in performance contain common spectral patterns that are shared across the dataset and the optimization process works to align them. Hence, the approach is likely to provide the best increase in performance when dealing with datasets in which there are significant regions of the DOS that remain constant across the structures, but are rigidly shifted when using conventional energy references. Furthermore, models trained under this framework are better able to extrapolate to larger system sizes. More generally, this paper highlights the significance and the benefits of defining evaluation metrics in a manner that is consistent with the physics of the problem. For instance, the concept of optimizing an internal or ill defined reference can be readily applied to applications outside of the DOS. To give a few examples, this approach can also be easily extended to the optimization of the energy reference for the prediction of the spatially resolved density of states. Additionally, it can also be applied to the prediction of the Hamiltonian of bulk structures \cite{zhong_universal_2024, li_deep-learning_2021, nigam_equivariant_2022, cignoni_electronic_2024}, whereby the eigenvalues obtained from the Hamiltonian do not have a well defined external energy reference. 

\section*{Supplemental Material}

The Supplemental Material contains further information on the model architecture and the reference datasets, as well as further benchmarks, and references to several relevant publications \cite{langreth_density_2009, rappe_optimized_1990, thonhauser_van_2007, perdew_restoring_2008, thonhauser_spin_2015, perdew_generalized_1996, marzari_thermal_1999, berland_van_2015}.

\section*{Data Availability}
The data that support the findings of this article are openly available on the Materials Cloud Archive \cite{how_adaptive_2024, talirz_materials_2020}. The code used to train the ML-DOS models can be found in the doslearn subpackage of the rholearn package on Github \cite{abbott_2024_13891847}.
An example of usage of these packages is also available at \url{https://atomistic-cookbook.org/examples/dos-align/dos-align.html}.

\begin{acknowledgments}
The authors thank Chiheb Ben Mahmoud for support during the early implementation of the method.
MC and SC acknowledge support by the Swiss National Science Foundation (Project No.  200020\_214879). WBH and MC acknowledge support by the MARVEL National Centre of Competence in Research (NCCR), funded by the Swiss National Science Foundation (Grant Agreement No. 51NF40-182892). FG~and MC~acknowledge funding from the European Research Council (ERC) under the European Union’s Horizon 2020 research and innovation program Grant No.~101001890-FIAMMA. FG~also acknowledges funding from the European Union's Horizon 2020 research and innovation program under the Marie Sk\l{}odowska-Curie Action IF-EF-ST, Grant Agreement No.~101018557-TRANQUIL.
\end{acknowledgments}
\appendix

\end{document}